\documentclass{article}
\usepackage{amsmath,amssymb}
\usepackage{graphicx}%
\newcommand{\Fig}[2]{%
\begin{center}
\parbox{8cm}{%
\refstepcounter{figure}\includegraphics[width=8cm]{#1} \noindent Figure \thefigure:\quad
#2}\end{center}}

\begin{document}
\begin{center}
{\bf \Large The Qualitative and Numerical Analysis of the Cosmological Model Based on Phantom Scalar Field with Self} \\[12pt]
Yu.G. Ignat'ev and A.A. Agathonov\\
N.I. Lobachevsky Institute of Mathematics and Mechanics, Kazan Federal University, \\ Kremleovskaya str., 35, Kazan, 420008, Russia
\end{center}

\begin{abstract}
In this paper we investigate the asymptotic behavior of the cosmological model based on phantom scalar field on the ground of qualitative analysis of the system of the cosmological model's differential equations and show that  as opposed to models with classical scalar field, such models have stable asymptotic solutions with constant value of the potential both in infinite past and infinite future. We also develop numerical models of the cosmological evolution models with phantom scalar field in this paper.

{\bf keywords}: cosmological model, phantom scalar field, quality analysis, asymptotic behavior, numerical simulation, numerical gravitation.\\
{\bf PACS}: 04.20.Cv, 98.80.Cq, 96.50.S  52.27.Ny
\end{abstract}

This work was funded by the subsidy allocated to Kazan Federal University for the state assignment in the sphere of scientific activities.

\section{The Introduction}

Previously, there was created the cosmological model based on the statistical systems of scalar charged particle with interparticle phantom scalar interaction possessing negative kinetic energy of field [1] – [8].  On the basis of the stated mathematical model there was carried out numerical simulation of both degenerated Fermi systems and charge-symmetrical Boltzmann plasma consisting of scalar charged particles and antiparticles [9] – [13]. These investigations revealed unique properties of the cosmological models which are based on the statistical systems of scalar charged par\-ticles with phantom scalar interaction. However, since general results were obtained using numerical simulation methods, they hardly can be used for description of asymptotic properties of the  corresponding models. Using combination of methods of the qualitative theory of ordinary differential equations and  numerical integration of them, in [14]-[15] there were investigated the asymptotic properties of the standard cosmological model based on classic massive scalar field. Parti\-cu\-larly, in these papers it is shown that the system of Einstein-Klein-Gordon equations for the homo\-ge\-nous space flat cosmological model has a single singular point corresponding to zero values of the scsalar field's potential and its derivative, where mentioned singular point can also be an attractive center or attractive focus, or attractive saddle. Besides, it was revealed a microscopic oscillating character of the invariant cosmological acceleration at approaching to the singular point with an average value corresponding to non-relativistic equation of state. In this paper we carry out similar investigation for the «standard» cosmological model based on phantom fields. In this model, as opposed to those, considered in [1] – [13], we do not account contribution of the ordinary matter, i.e. we consider free phantom fields without a source.

\section{Basic Relations for the Cosmological Model with a Phantom Scalar Field}
\subsection{Equations of Free Scalar Field with Self-Action}
The Lagrange function of the phantom scalar field with mass $m$ and self-action has a form [3]:
\begin{equation} \label{GrindEQ__1_}
L=-\frac{1}{8\pi } \left(g^{ik} \Phi _{,i} \Phi _{,k} +m^{2} \Phi ^{2} +\frac{\alpha }{2} \Phi ^{4} \right),
\end{equation}
where $\alpha $ is a constant of self-action. Tensor of the energy-momentum relative to this function
\begin{eqnarray} \label{GrindEQ__2_}
T^{ik} =\frac{1}{8\pi } \left(-2\Phi ^{,i} \Phi ^{,k} +g^{ik} \Phi _{,j} \Phi ^{,j}\right.\nonumber\\
 \left.+g^{ik} m^{2} \Phi ^{2} +g^{ik} \frac{\alpha }{2} \Phi ^{4} \right)
\end{eqnarray}
has a negative kinetic term. If covariant divergence of this tensor is zero, then we have the following equation of free phantom scalar field:
\begin{equation} \label{GrindEQ__3_}
\square \Phi -m_{^{*} }^{2} \Phi =0,
\end{equation}
where
\begin{equation} \label{GrindEQ__4_}
m_{*}^{2} =m^{2} +\alpha \Phi ^{2}
\end{equation}
is an effective mass of a scalar boson. The equation \eqref{GrindEQ__3_} is different from the equation of Klein-Gordon in the fact of presence of cubic nonlinearity and negative sign of the massive term.
Let us write out the Einstein equations with the cosmological term $\Lambda >0$ \footnote{We use the Planck system of units:  $G=c=\hbar =1$ . }
\begin{equation} \label{GrindEQ__5_}
R^{ik} -\frac{1}{2} Rg^{ik} =\Lambda g^{ik} +8\pi T^{ik} .
\end{equation}

\subsection{The Self-Consistent Equations for the Space-Flat Friedmann Model}
Let us write out the self-consistent equations of the space-flat cosmological model
\begin{equation} \label{GrindEQ__6_}
ds^{2} =dt^{2} -a^{2} (t)(dx^{2} +dy^{2} +dz^{2})
\end{equation},
the Einstein equation
\begin{equation} \label{GrindEQ__7_}
3\frac{\dot{a}^{2} }{a^{2} } =-\dot{\Phi }^{2} +m^{2} \Phi ^{2} +\frac{\alpha }{2} \Phi ^{4} +\Lambda
\end{equation}
and the equation of massive phantom scalar field with cubic nonlinearity \footnote{Here and further it is $\dot{f}\equiv df/dt$.}:
\begin{equation} \label{GrindEQ__8_}
\ddot{\Phi }+3\frac{\dot{a}}{a} \dot{\Phi }-m_{*}^{2} \Phi =0.
\end{equation}
Herewith the energy-momentum tensor \eqref{GrindEQ__2_} has a structure of energy -- momentum tensor of ideal flux with the following energy density and pressure:
\begin{eqnarray} \label{GrindEQ__9_}
\varepsilon =\frac{1}{8\pi } \left(-\dot{\Phi }^{2} +m^{2} \Phi ^{2} +\frac{\alpha }{2} \Phi ^{4} \right);\nonumber\\
p=-\frac{1}{8\pi } \left(\dot{\Phi }^{2} +m^{2} \Phi ^{2} +\frac{\alpha }{2} \Phi ^{4} \right),
\end{eqnarray}
such that:
\[\varepsilon +p=-\frac{\dot{\Phi }^{2} }{4\pi } .\]

\subsection{The Kinematic Invariants}
Further we will need also the values of two kinematic invariants of the Friedmann Universe:
\begin{equation} \label{GrindEQ__10_}
H(t)=\frac{\dot{a}}{a} \ge 0;{\rm \; \; }\Omega (t)=\frac{a\ddot{a}}{\dot{a}^{2} } \equiv 1+\frac{\dot{H}}{H^{2} }
\end{equation}
-- the Hubble constant and the invariant cosmological acceleration.

\section{The Qualitative Analysis}
\subsection{Reducing the System of Equations to Canonical Form}
Using the fact that the Hubble constant can be expressed through the Einstein equation  \eqref{GrindEQ__7_} in functions $\Phi ,{\rm \; }\dot{\Phi }$, proceeding to dimensionless Compton time:
\[mt=\tau;\quad (m\not\equiv0) \]
and carrying out standard change of variables $\Phi '=Z(t)$, let us reduce the field equation \eqref{GrindEQ__8_} to the form of normal autonomous system of ordinary differential equations in the plane $\{ \Phi ,Z\} $:
\begin{eqnarray} \label{GrindEQ__11_}
\Phi '=Z;& Z'=\displaystyle-\sqrt{3} \sqrt{\Lambda _{m} -Z^{2} +\Phi ^{2} +\frac{\alpha _{m} }{2} \Phi ^{4} } Z\nonumber\\
& \displaystyle +\Phi +\alpha _{m} \Phi ^{3} ,
\end{eqnarray}
where $f'\equiv df/d\tau $ and the following denotations are introduced:

\[\Lambda _{m} \equiv \frac{\Lambda }{m^{2} } ;{\rm \; \; }\alpha _{m} \equiv \frac{\alpha }{m^{2} } . \]
Herewith it is:

\begin{equation} \label{GrindEQ__12_}
H=m\frac{a'}{a} \equiv mh;{\rm \; \; \; }\Omega =\frac{aa''}{a'^{2} } \equiv 1+\frac{h'}{h^{2} } .
\end{equation}

Thus, we have an autonomous two-dimensional dynamic system in the phase plane $\{ \Phi ,Z\} $. To reduce it to standard denotations of the qualitative theory of differential equations (see e.g.,  [16]) let us put:
\begin{eqnarray}\label{GrindEQ__13_}
\Phi =x;\;Z=y; P(x,y)=y;\; Q(x,y)=\nonumber\\
-\sqrt{3}\sqrt{\Lambda_{m}-y^{2} +x^{2}+\frac{\alpha_{m}}{2} x^{4}}y+x+\alpha_{m}x^{3} .
\end{eqnarray}
The corresponding normal system of equations in standard denotations has the following form:
\begin{equation} \label{GrindEQ__14_}
x'=P(x,y);{\rm \; \; \; }y'=Q(x,y) .
\end{equation}
To have a real solution of the system of differential equations \eqref{GrindEQ__11_} (or (14)), it is required that the following inequality is fulfilled:
\begin{equation} \label{GrindEQ__15_}
\Lambda _{m} -y^{2} +x^{2} +\frac{\alpha _{m} }{2} x^{4} \ge 0.
\end{equation}

\subsection{The Singular Points of the Dynamic System}
The singular points of the dynamic system are defined through equations (see e.g.  [16]) :
\begin{equation} \label{GrindEQ__16_}
M:{\rm \; \; }P(x,y)=0;{\rm \; }Q(x,y)=0.
\end{equation}
It is apparent, that at any $\alpha _{m} $ è $\Lambda _{m} \ge 0$ the system of algebraic equations \eqref{GrindEQ__15_} as well as in papers [14]-[15] has the single trivial solution
\begin{equation} \label{GrindEQ__17_}
x=0;y=0{\rm \; }\Rightarrow M_{0} (0,0) .
\end{equation}
Besides, in the case $\alpha _{m} <0$  the following non-trivial symmetrical solutions are possible:
\begin{equation} \label{GrindEQ__18_}
x=x_\pm=\pm \frac{1}{\sqrt{-\alpha_{m}}};y=0\quad \Rightarrow M_\pm(x_\pm,0).
\end{equation}
Substituting the solutions \eqref{GrindEQ__18_} in the condition \eqref{GrindEQ__15_}, we find the necessary condition of the solutions in the singular points  \eqref{GrindEQ__17_} and \eqref{GrindEQ__18_} to be of real type:
\begin{equation} \label{GrindEQ__19_}
(17)\rightarrow \Lambda_m\ge 0;(18)\rightarrow\Lambda_m -\frac{1}{2\alpha_m}\ge 0.
\end{equation}
The second of these conditions is always fulfilled as a consequence of the first one at $\alpha _{m} <0$.

\subsection{The Characteristic Equation and the Qualitative Analysis in the Case Near Zero Singular Point}
Let us calculate the derivatives of functions \eqref{GrindEQ__13_} in zero singular point \eqref{GrindEQ__16_} at $\Lambda _{m} >0$ :

\begin{eqnarray}
\left. \frac{\partial P}{\partial x} \right|_{M_{0} } =0; & \displaystyle \left. \frac{\partial P}{\partial y} \right|_{M_{0} } &=1; \nonumber\\
\left. \frac{\partial Q}{\partial x} \right|_{M_{0} } =1;& \displaystyle\left. \frac{\partial P}{\partial y} \right|_{M_{0} } &=-\sqrt{3\Lambda_{m}}. \nonumber
\end{eqnarray}
Thus, we find the following characteristic equation (see [16]):
\begin{eqnarray} \label{GrindEQ__19_}
\left|
\begin{array}{cc}
{-\lambda } & {1} \\[12pt]
{1} & {-\lambda -\sqrt{3\Lambda _{m} } }
\end{array}
\right|=0\nonumber\\[12pt]
\Rightarrow \lambda _{\pm } =-\frac{\sqrt{3\Lambda _{m} } }{2} \pm \frac{\sqrt{3\Lambda _{m} +4} }{2} ,
\end{eqnarray}

\noindent so that:

\begin{equation} \label{GrindEQ__20_}
\lambda _{+} \lambda _{-} =-1
\end{equation}
both eigenvalues are real values having opposite signs. It is not difficult to see that eigenvectors of $\lambda $ - matrix are orthogonal:
\begin{equation} \label{GrindEQ__21_}
\mathbf{u}_\pm =(\lambda_\pm ,1)\Rightarrow (\mathbf{u}_{+,}\mathbf{u}_{-} )=0.
\end{equation}
 Thus, according to the qualitative theory of the dynamic systems in the plane (see e.g. [16]) the considered dynamic system  \eqref{GrindEQ__11_} in the case $\Lambda_m>0$ always has a \textit{saddle singular point}\eqref{GrindEQ__16_}, which is \textit{instable at any direction of time}. In the case $\alpha>0$ this point is a unique singular point of the dynamic system. The zero singular point has 4 orthogonal separatrices which correspond to exceptional phase trajectories coming into this point at $t\to \pm \infty $ along the directions of eigenvectors \eqref{GrindEQ__21_} (Fig.1). This means that, first of all, phase trajectories do not come into a singular point at $t\to \pm \infty$ in any cases apart from some exceptional ones, i.e. near this point it is either $\Phi \ne 0$, or  $Z\ne 0$.

 Here it is seen that the entire phase plane is separated into 4 squares where in left and right (symmetrical) quadrants it is $|\Phi |>|Z|$ everywhere while in top and bottom (symmetrical) quadrants it is everywhere $|\Phi |<|Z|$. Thus, when non-exceptional phase trajectories approach the singular point, it is either a). $Z\to 0;\;\Phi\to\Phi_0\ne 0$, or  b). $\Phi\to 0;\;Z\to Z_0 \ne 0$.

\Fig{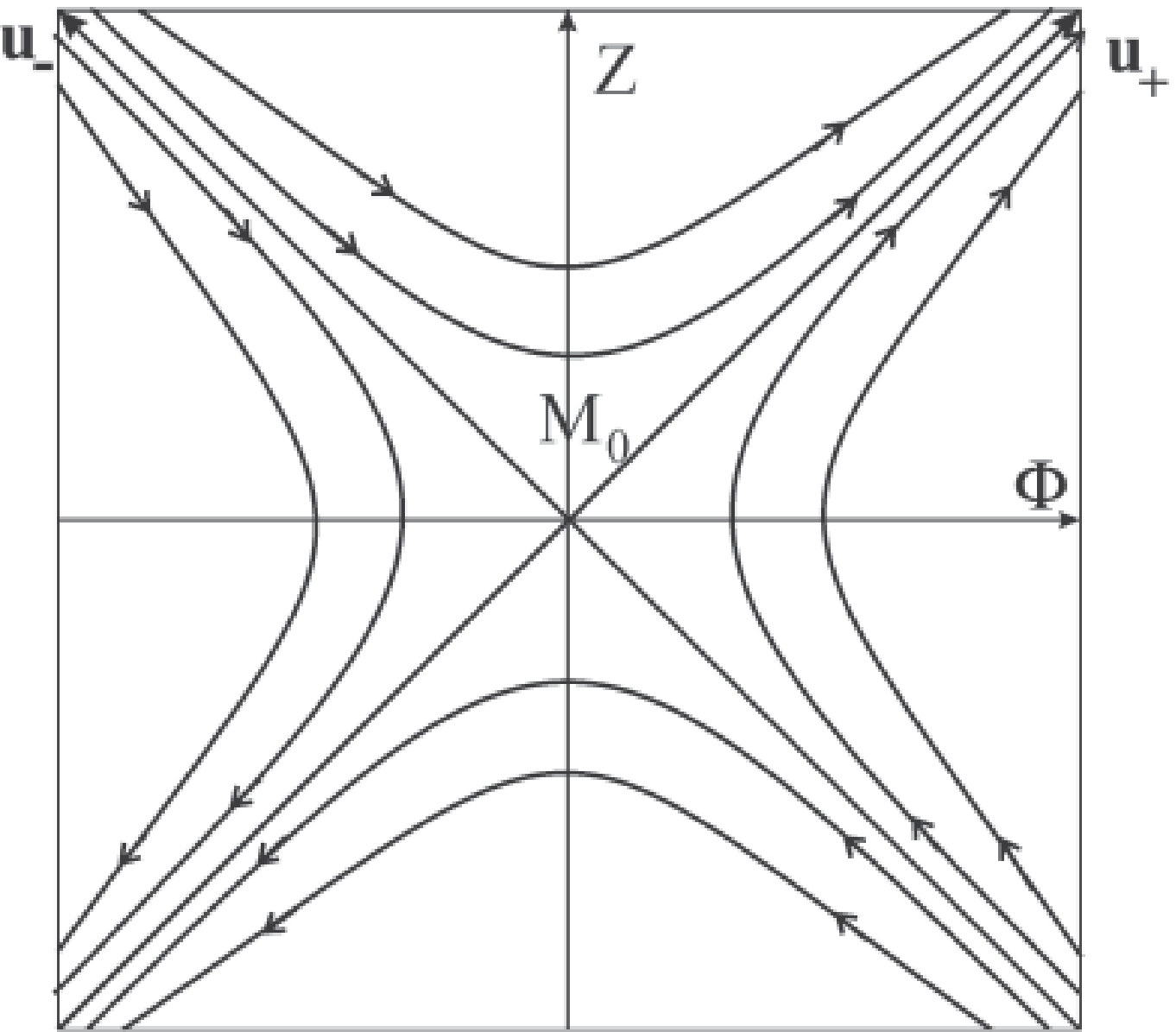}{The phase portrait of the system \eqref{GrindEQ__11_} at $\alpha >0$. An instable saddle.}

\subsection{Numerical Simulation for $\alpha=0$}
Numerical simulation of the dynamic system is carried out in applied mathematics package Maple. On the plots presented below there are shown certain results of numerical integration of the equations  \eqref{GrindEQ__11_} using Rosenbrock method. These results demonstrate the most characteristic properties of the cosmological system's dynamics (Fig. 2 - 4).
\Fig{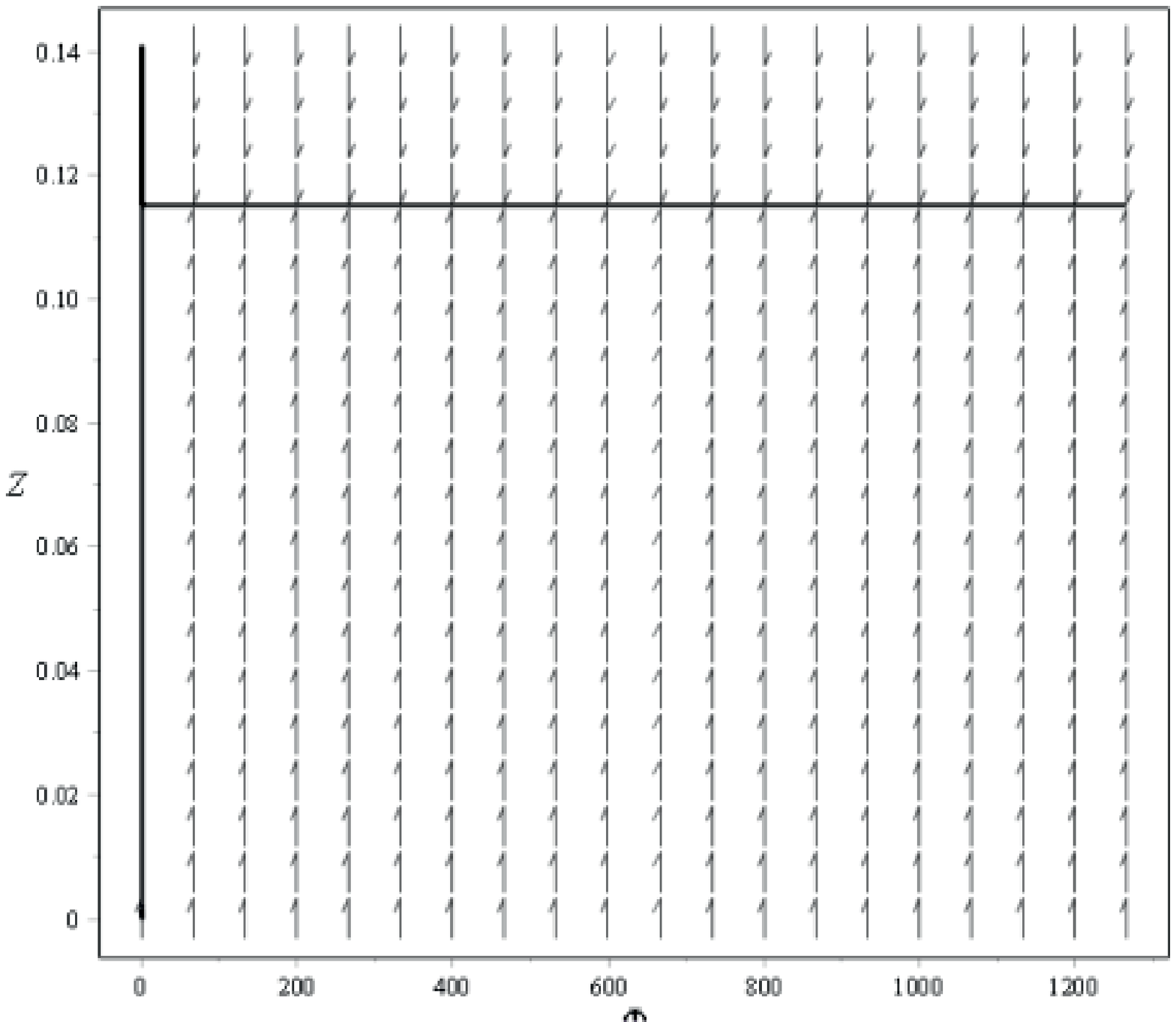}{The large-scale picture of the system's  \eqref{GrindEQ__11_} phase portrait for the case $\alpha=0$; $\Lambda_m=0.0001$; $\Phi(-1000)=0.001$; $Z(-1000)=0$; $t=-1000\div 10000$.}

On the presented phase portraits of the dynamic system (\ref{GrindEQ__11_}), obtained by numerical integration, we can trace the basic properties of the cosmological evolution. At early stage one can observe a rapid growth of the potential's derivative $\Phi'=Z$, then there follows a long evolution part where it is $Z\approx \mathrm{Const}$ and the value of the scalar potential grows linearly with time.

Thus, represented on Fig. 2 -- 4 phase portraits numerically realize one of the phase trajectories shown on qualitative diagram on Fig. 1, namely the trajectory in the right part of the diagram.

\Fig{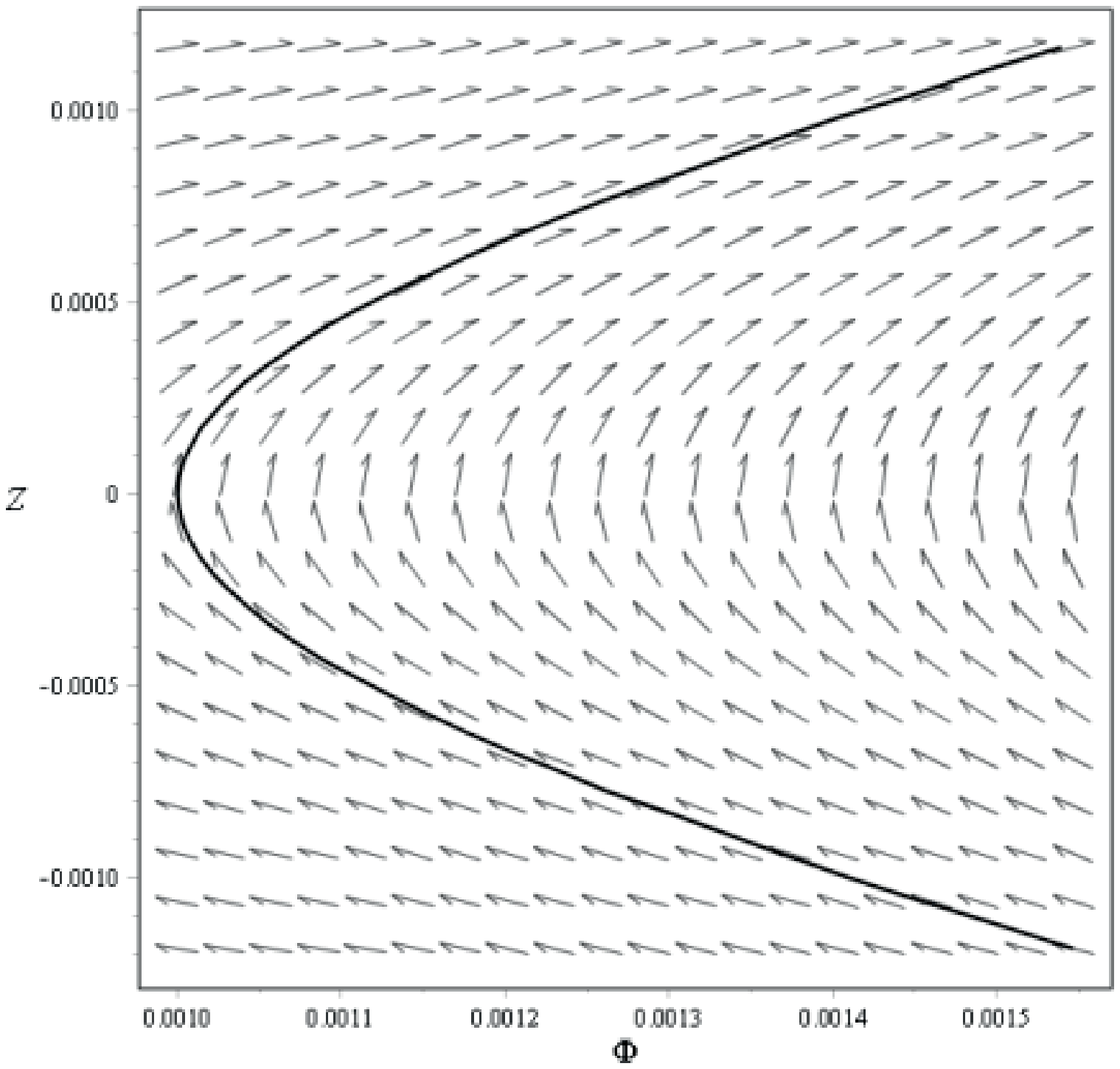}{The initial part of the system's \eqref{GrindEQ__11_} phase portrait for the case $\alpha=0$; $\Lambda_m=0.0001$; $\Phi(-1000)=0.001$; $Z(-1000)=0$; $t=-1001\div -999$.}
\Fig{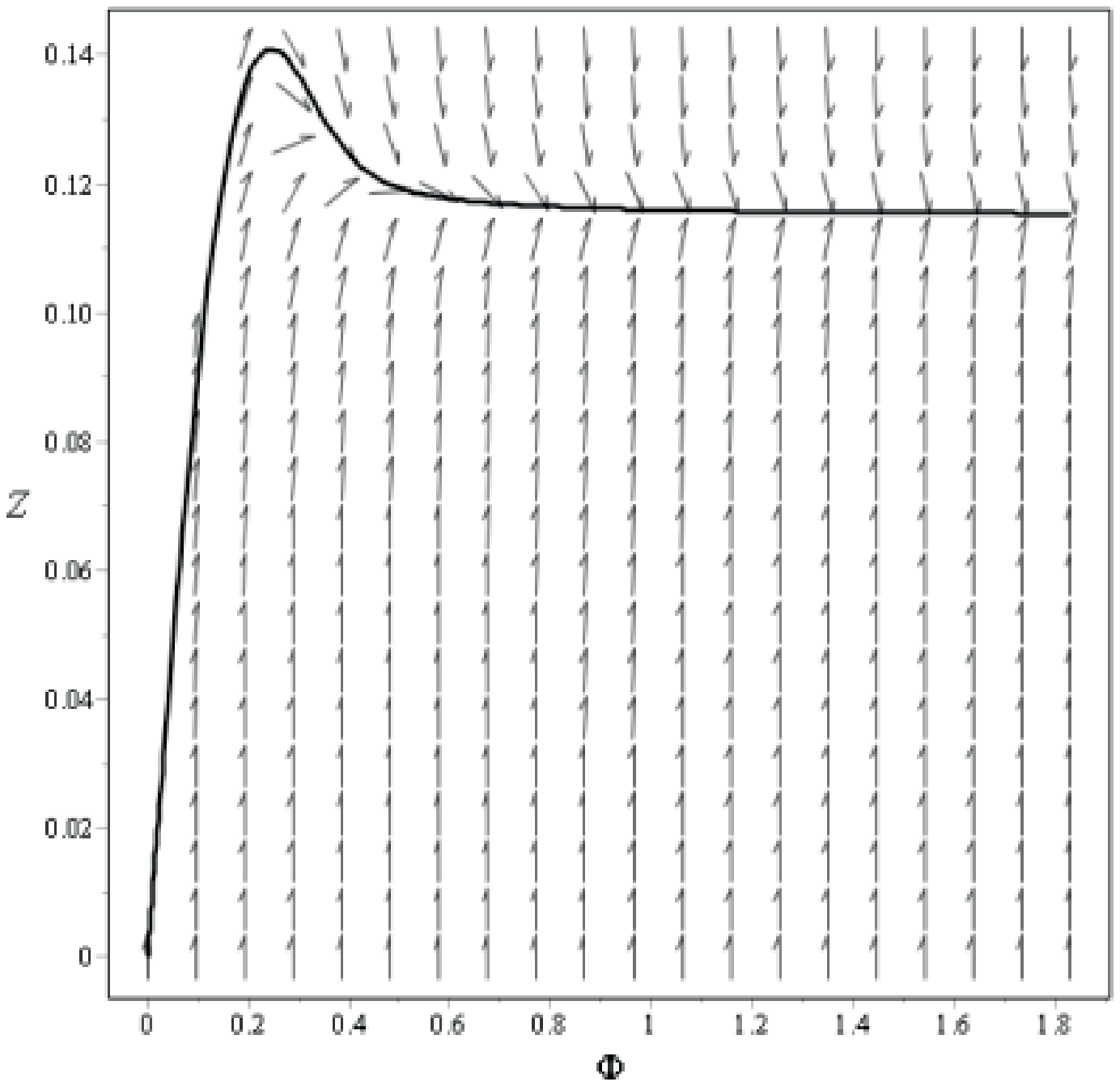}{The derivative's burst at the initial part of the system's \eqref{GrindEQ__11_} phase portrait for the case $\alpha=0$; $\Lambda_m=0.0001$; $\Phi(-1000)=0.001$; $Z(-1000)=0$; $t=-1000\div -980$.}

\subsection{The Characteristic Equation and Qualitative Analysis in the Case $\alpha _{m}<0$}
The derivative of functions \eqref{GrindEQ__13_} in singular points \eqref{GrindEQ__18_} at $\Lambda _{m} >0$ are equal to:
\begin{eqnarray}
\left.\frac{\partial P}{\partial x} \right|_{M_\pm=0;} &\displaystyle \left.\frac{\partial P}{\partial y}\right|_{M_\pm =1};\nonumber \\
\left.\frac{\partial Q}{\partial x} \right|_{M_\pm =-2};& \left. \displaystyle\frac{\partial P}{\partial y} \right|_{M_\pm} =-\displaystyle
\sqrt{3}\sqrt{\Lambda _{m}-\frac{1}{2\alpha _{m}}}.\nonumber
\end{eqnarray}
The characteristic equations for both singular points coincide with each other and singular points have the single type (see [16]):
\begin{eqnarray} \label{GrindEQ__22_}
\left|\begin{array}{cc}
-\lambda  & 1 \\
1 &\displaystyle -\lambda -\sqrt{3}\sqrt{\Lambda_m -\frac{1}{2\alpha_m}}\\[12pt]
\end{array}\right|=0\nonumber\\
\Rightarrow \lambda_\pm =-\frac{\sqrt{3}\sqrt{\Lambda _m-\displaystyle\frac{1}{2\alpha_m}}}{2} \nonumber\\
\pm \frac{\sqrt{3}\sqrt{\Lambda_m-\displaystyle\frac{1}{2\alpha_m}-\frac{8}{3}}}{2}.
\end{eqnarray}
In consequence of \eqref{GrindEQ__19_} the radicand in the first term \eqref{GrindEQ__22_} is strictly greater than zero therefore the following three cases are possible:\\
1) $\Lambda_m-1/2\alpha_m-8/3>0$ -- then two eigenvalues are real and negative. In this case the solution contains \textit{two symmetrical attractive (stable) non-degenerated nodes}. All the phase trajectories around such singular points at $t\to\infty$ come into these points and, apart from two exceptional ones, are tangent to the eigenvector of the minimal length.
The eigenvectors are again defined by means of formula (\ref{GrindEQ__21_}), where it is necessary to substitute corresponding eigenvalues from (\ref{GrindEQ__22_}). It is not difficult to show that $\mathbf{u}_+$ (Fig. 5) is this eigenvector of the minimal length.\\
2) $\Lambda_m-1/2\alpha_m-8/3=0$ -- then both eigenvalues are negative and equal to each other. In this case the solution contains \textit{two symmetrical degenerated nodes}. The phase portrait practically coincide with the portrait on Fig.5.\\
3) $\Lambda_m-1/2\alpha_m-8/3<0$ -- then both eigenvalues are complex conjugated and their real parts are negative. In this case the solution contains\textit{ two symmetrical attractive focuses} (Fig. 6).
\Fig{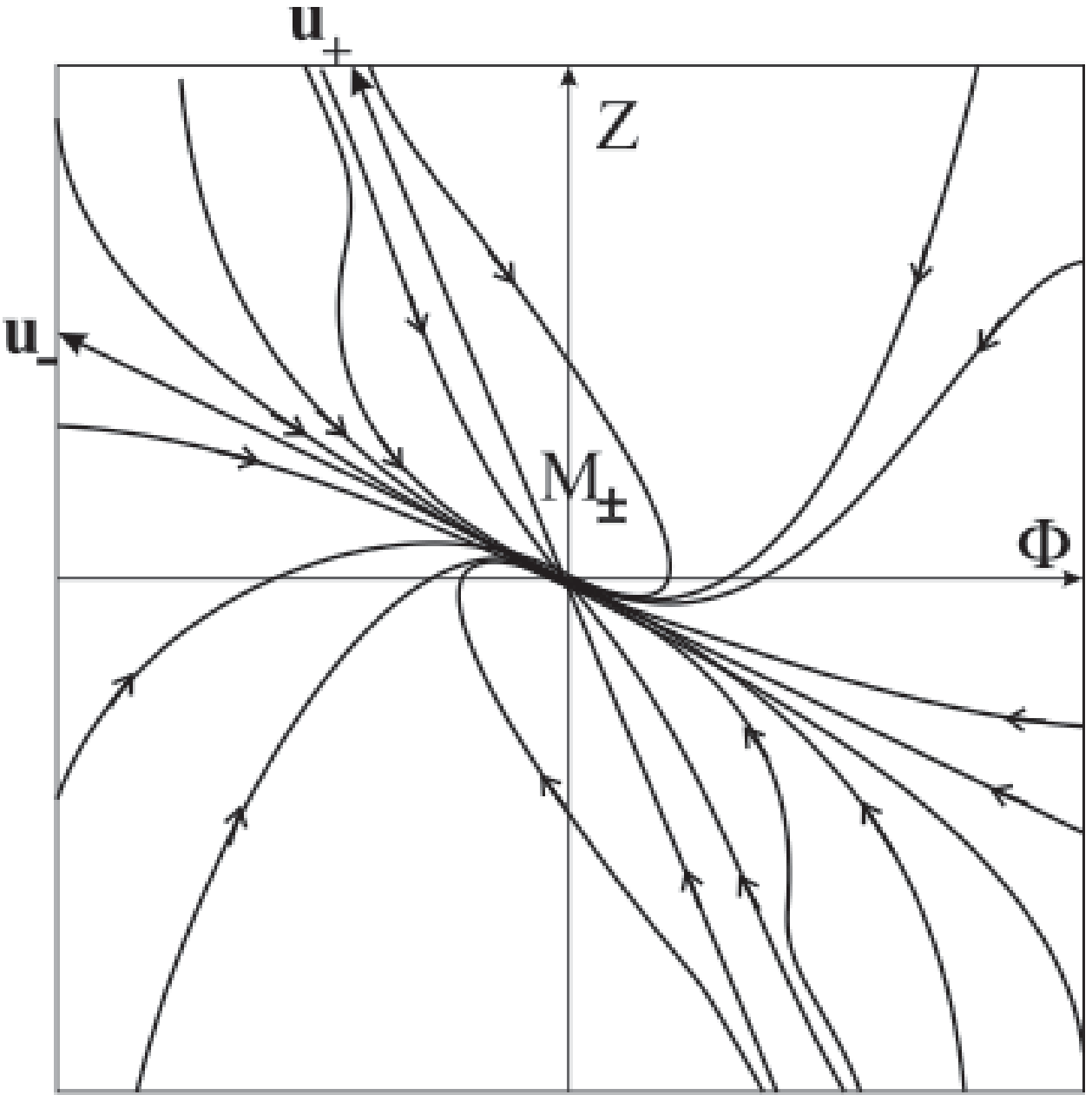}{The phase portrait of the system \eqref{GrindEQ__11_} at $\alpha <0$; $\Lambda_m-1/2\alpha_m-8/3>0$. One of the symmetrical attractive stable nodes.}
\Fig{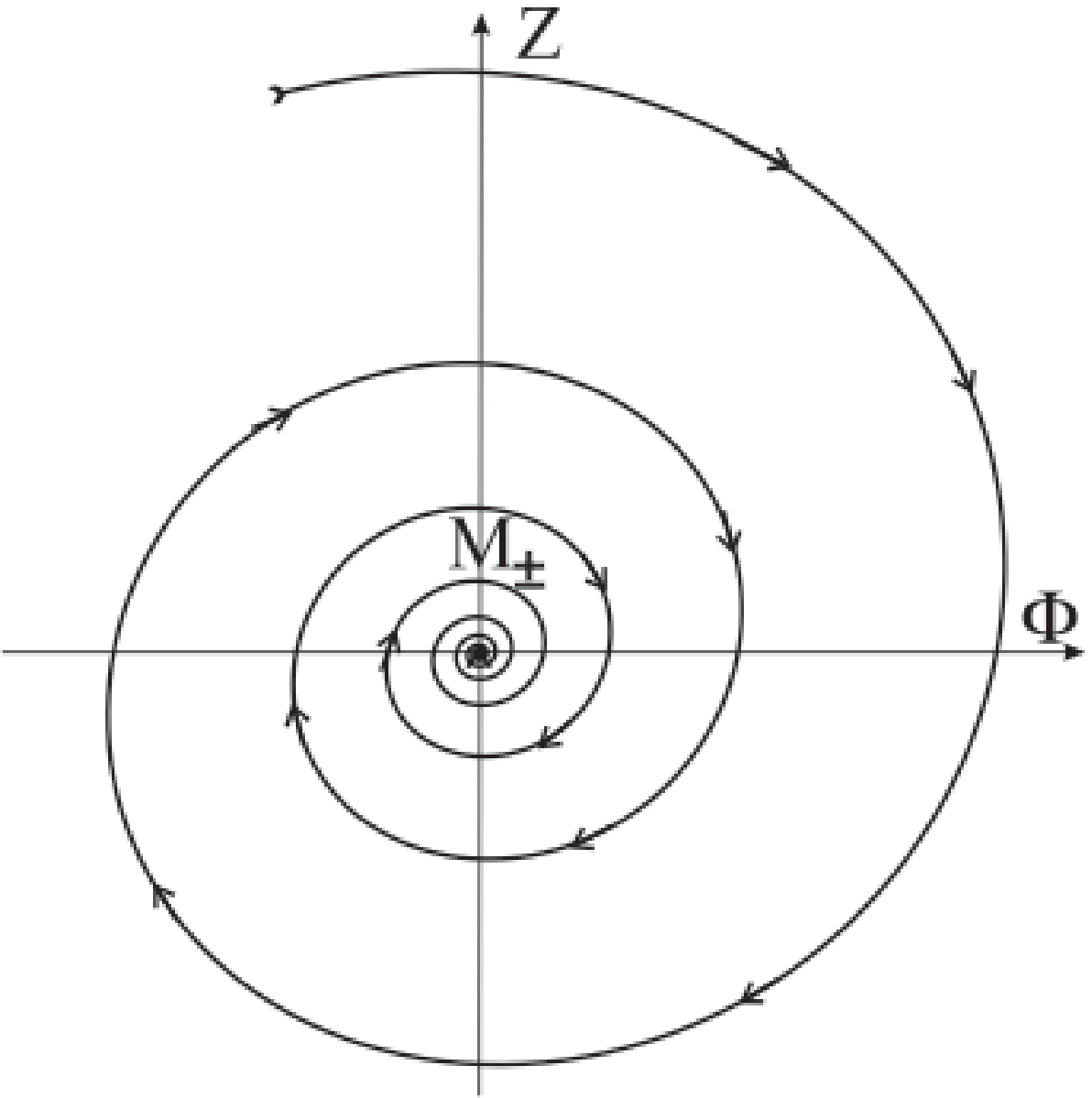}{The phase portrait of the system \eqref{GrindEQ__11_} at $\alpha <0$; $\Lambda_m-1/2\alpha_m-8/3<0$. One of the symmetrical attractive stable focuses $\mathrm{Re}(\lambda)<0$ -- clockwise rotation.}
\subsection{Numerical Simulation for $\alpha<0$}
The numerical simulation of the dynamic system was carried out in the applied mathematics package Mathematica. On the presented below plots (7 -- 12) there are shown the results of the numerical equations \eqref{GrindEQ__11_}, demonstrating the most characteristic properties of the cosmological system's dynamics.
\Fig{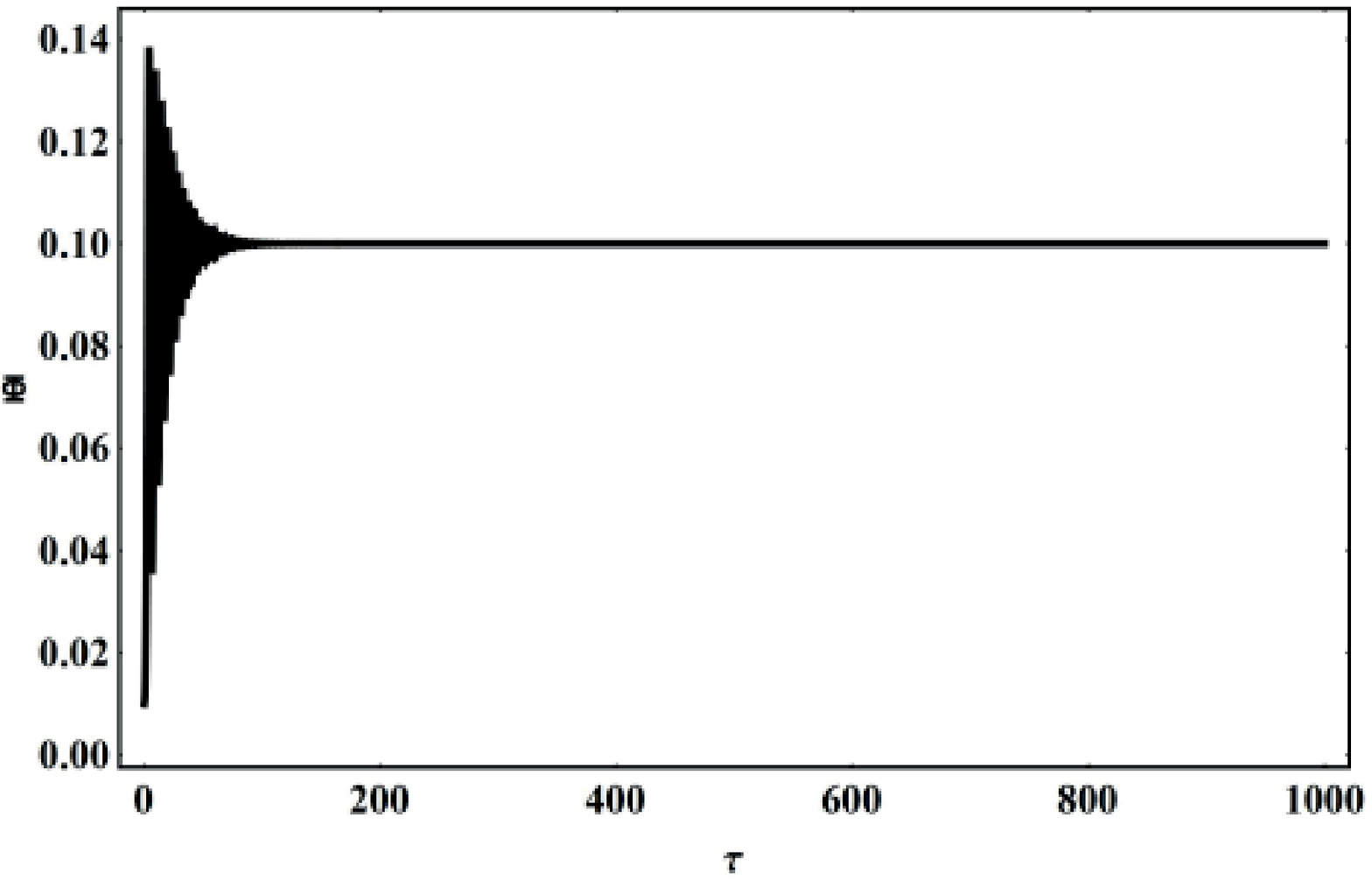}{The cosmological evolution of the potential $\Phi(\tau)$ at $\alpha_m=-100$; $\Lambda_m=0.00001$; $\Lambda_m-1/2\alpha_m-8/3=-2.661656667$; $\Phi(0)=0.01$, $Z(0)=0$.}
\Fig{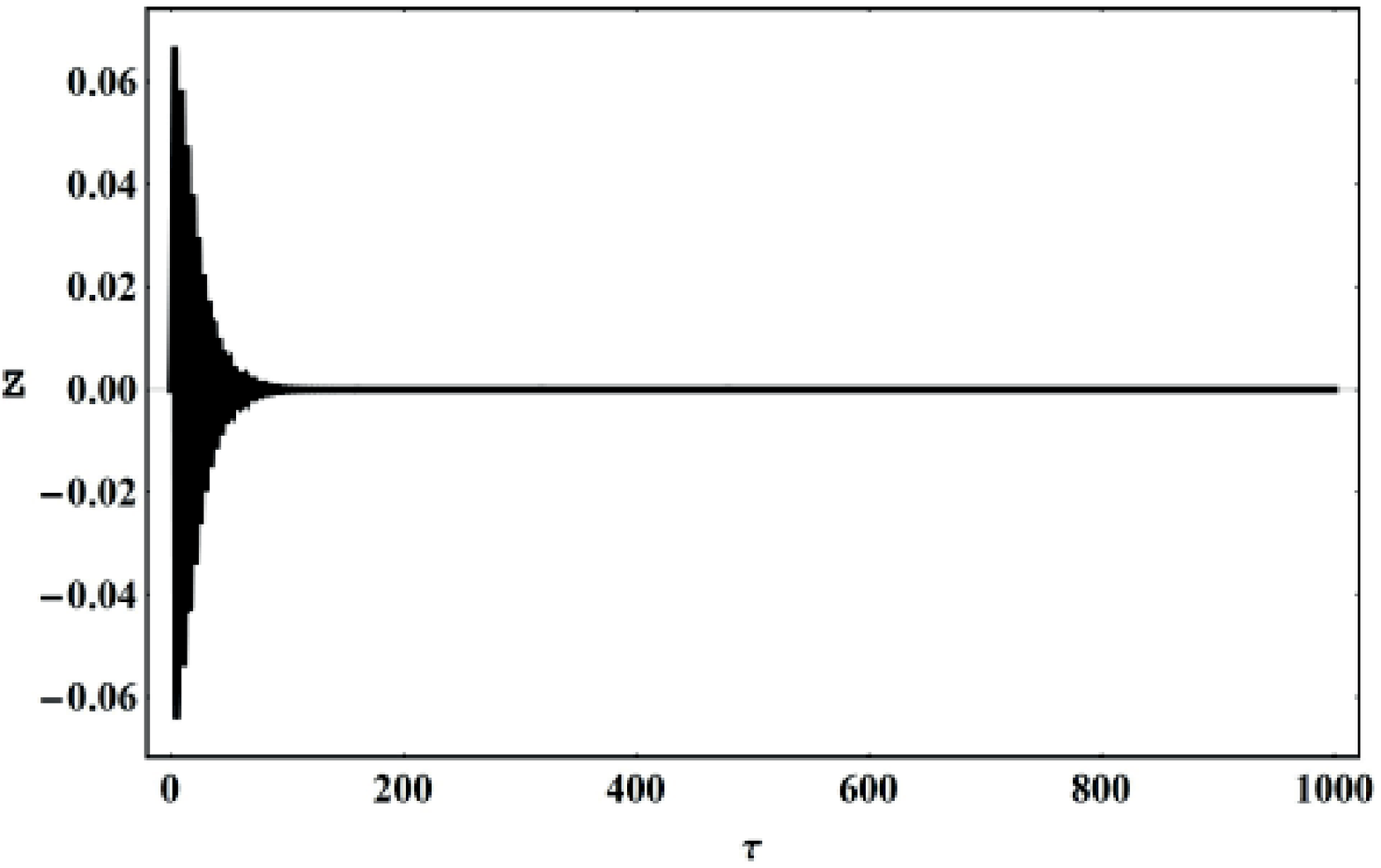}{The cosmological evolution of the potential's derivative $Z(\tau)$ for the same case.}
\Fig{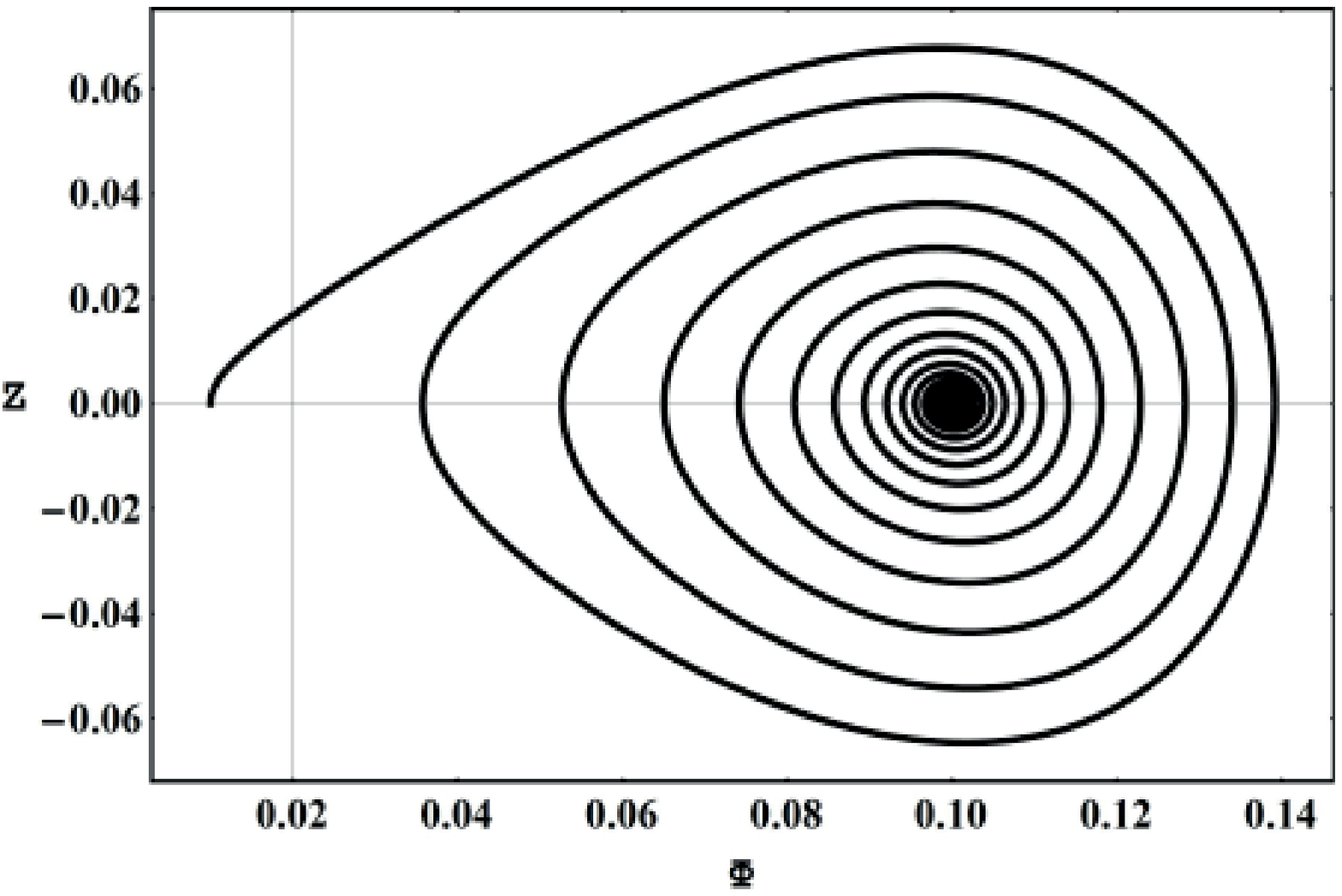}{The part of the system's \eqref{GrindEQ__11_} phase portrait for the same case. The right attractive stable focus $\mathrm{Re}(\lambda)>0$ -- rotation counterclockwise. The focus corresponds to $\Phi\approx 0.1$;
$\tau=0\div1000$.}
\Fig{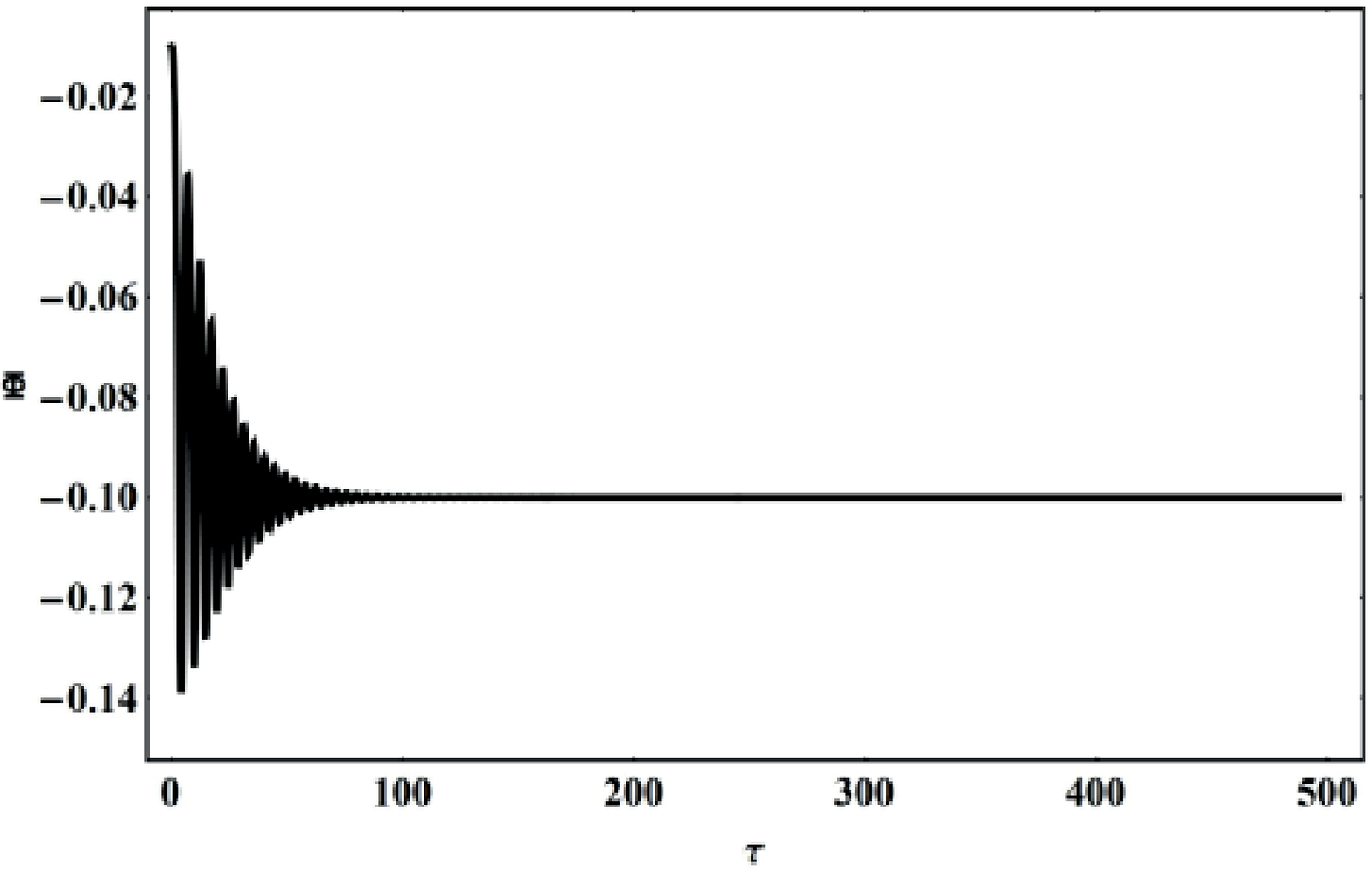}{The cosmological evolution of the potential $\Phi(\tau)$ at $\alpha_m=-100$; $\Lambda_m=0.00001$; $\Lambda_m-1/2\alpha_m-8/3=-2.661656667$; $\Phi(0)=-0.01$, $Z(0)=0$.}
\Fig{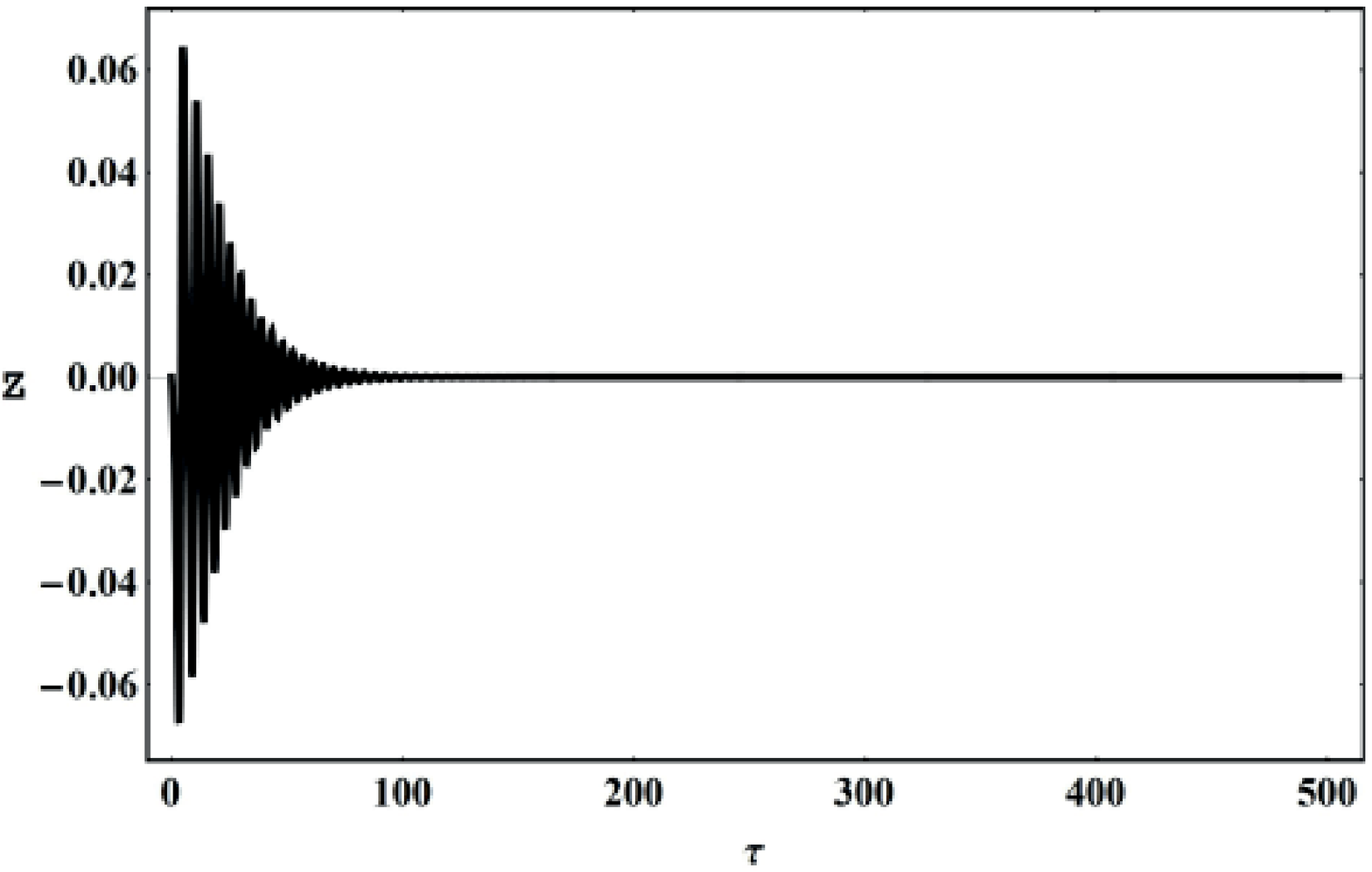}{The cosmological evolution of the potential's derivative $Z(\tau)$ for the same case.}
\Fig{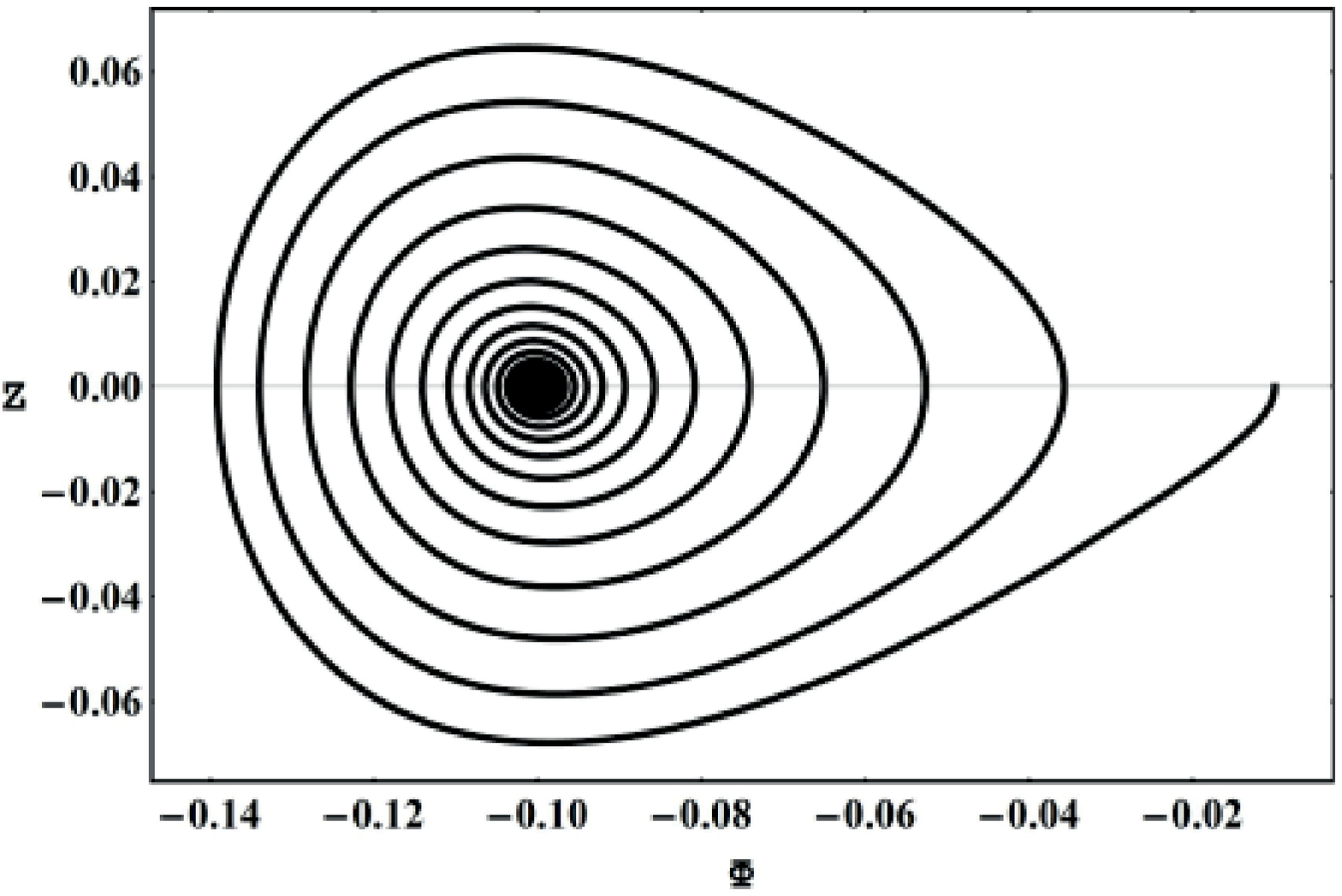}{A part of the phase portrait of the system \eqref{GrindEQ__11_} for the same case. The left attractive stable focus $\mathrm{Re}(\lambda)>0$ -- rotation counterclockwise. Focus corresponds to $\Phi\approx -0.1$;
$\tau=0\div1000$.}

\section{The Conclusion}

It is not difficult to see that the system \eqref{GrindEQ__11_} is invariant relative to transformation
\begin{equation} \label{GrindEQ__22_}
\Phi \to -\Phi ;{\rm \; Z}\to {\rm -Z},
\end{equation}
i.e. with respect to reflection from the zero singular point however it is not invariant with respect to reflection from the coordinate axes. Moreover, it is seen that at instants of time  $t=\pm \infty $ variables $|\Phi (\pm \infty )|\to \infty {\rm \; }$; ${\rm Z(}\pm \infty )|\to \infty $ , i.e. in this model infinite values of the scalar potential and its derivative are reached both in infinite past and infinite future. Herewith phase trajectories tend to separatrices: $Z=\pm \Phi $, i.e., $\Phi \sim e^{\pm t} $. Therefore at $\alpha>0$ in the infinite past and infinite future it is $H=\mathrm{Const}\Rightarrow \Omega =1$, i.e. both in the infinite past and infinite future the Universe is situated at the inflation stage.

In the case of negative constant of interaction in the most interesting case when there exist two symmetrical focuses, the solution of the equations of the cosmological model tends to inflation one in the infinite future and in contrast to standard model with classical scalar field it is $\Phi(+\infty)=\Phi_\infty\not=0$ and $\Phi(+\infty)\not= \Phi(-\infty)$. In this case the late inflation can be maintained both by cosmological constant and late scalar field which opens the way for manipulating the cosmological models to match them to observational data.

In conclusion, the Authors express their gratitude to the members of MW seminar for relativistic kinetics and cosmology of Kazan Federal University for helpful discussion of the work.



\begin{thebibliography}{}
%
\bibitem{1}  Yu. G. Ignat'ev, Russ. Phys. J.   {\bf 55}  166-172 (2012).
\bibitem{2}  Yu. G. Ignatiev, Russ. Phys. J. {\bf 55}  550-560 (2012).
\bibitem{3}  Yu. G. Ignatyev, Russ. Phys. J. {\bf 55} 1345-1350  (2012).
\bibitem{4}  Yu. G. Ignat'ev Yu. G.,  Space, Time and Fund. Interact. {\bf No 1}  47-69 (2014).
\bibitem{5}  Yu. G. Ignatyev Yu.G. and  D. Yu. Ignatyev,  Grav. and Cosmol. {\bf 20} 299–303  (2014).
\bibitem{6}  Yu. G. Ignatyev, Grav. and Cosmol. {\bf 21} 296–308 (2015).
\bibitem{7}  Yu. G. Ignatyev and  A. A. Agathonov,  Grav. and Cosmol. {\bf 21} 105–112 (2015).
\bibitem{8}  Yu. G. Ignat'ev, Space, Time and Fund. Interact. {\bf No 1} 79-98 (2012).
\bibitem{9}  Yu. G. Ignatyev, A. A. Agathonov and  D. Yu. Ignatyev, Grav. and Cosmol. {\bf 20} 304–308 (2014); // arXiv:1608.05020 [gr-qc].
\bibitem{10} Yu. G.Ignat'ev and M. L. Mikhailov, Russ. Phys. J. {\bf 57} 1743-1752. (2014).
\bibitem{11} Yurii Ignat'ev,  Alexander Agathonov,  Mikhail Mikhailov and  Dmitry Ignatyev,  Astroph. Space Sci  {\bf 357:61} (2015).
\bibitem{12} Yurii Ignat'ev,  Alexander Agathonov and Dmitry  Ignatyev,	arXiv:1608.05020 [gr-qc].
\bibitem{13} Yu. G. Ignatyev ,  M. L. Mikhailov, Space, Time and Fund. Interact. {\bf No 1} 75-90  (2015).
\bibitem{14} Yurii Ignat'ev // arXiv:1609.00745 [gr-qc].
\bibitem{15} Yu. G. Ignat'ev, Space, Time and Fund. Interact. {\bf No 3}  17-31 ( 2016).
\bibitem{16}  O. I. Bogoyavlensky, The methods of the qualitative theory of dynamic systems in astrophysics and gas dynamics. Moskow, Nauka (1980).
\end{thebibliography}
\end{document}